\documentstyle[12pt]{article}
\normalsize
\begin {document}
\overfullrule 0 pt
\null
\vskip 0 pt
\begin{center}
{\large \bf ON THE KINETIC PROPERTIES OF SOLITONS IN 
NONLINEAR SCHR\"{O}DINGER EQUATION} \\
\vskip 0.3 in
		           I.V.Baryakhtar  \\

N.N.Boholubov Institute for Theoretical Physics, vul. Metrologichna 14b, 
Kiev 252130, Ukraine \\
		   	    	and \\
Institute for Low Temperature Physics and Engineering, 47 Lenin Ave., Kharkov 
310164, Ukraine \\

	     	   	{\bf Abstract}     \\

\end{center}
The Boltzmann type kinetic equation for solitons in Nonlinear Schr\"{o}dinger 
equation has been constructed on the base of analysis of two soliton 
collision. Possible applications for Langmuir solitons in plasma and solitons 
in optic fibers are discussed.

\newpage

1. Nonlinear Schr\"{o}dinger equation (NSE) is one of the most popular 
equations for describing nonlinear phenomena in condensed matter and plasma. 
Such effects as small amplitude localized waves in magnets, localized 
excitations in quasi-one-dimensional biologic systems and optic fibers are 
described by NSE (see for example [1-3]).
In spite of wide field of applications, kinetic properties of solitons in NSE
as well as the another models were investigated much less then kinematic and 
dynamic properties of solitons. Some aspects of kinetic behavior of solitons 
in Korteveg-de-Vries, $\phi^4$ and Sine-Gordon equations have
been investigated in [4-8]. Fokker-Plank equation for the distribution 
function 
of soliton in NSE damping and fluctuations were considered in [9].
In present paper the kinetic equation for solitons in  NSE following the
approach suggested in [10] is proposed. Possible applications for solitons in 
plasma  and optic fibers are discussed.

2. NSE in dimensionless form for slow varying nonlinear envelope is  written 
as:                  		                           

\begin{equation}
i \frac {\partial \psi}{\partial t} + \frac {\partial^2 \psi} {\partial x^2}
+ 2 \psi |\psi|^2 =0.                                          \label{f1}
\end{equation}

Construction of kinetic equation for solitons needs  analysis of the 
process of soliton-soliton interaction. It is well known that solitons in  
exactly integrable models do not change there forms and velocities and 
additional shifts of there coordinates $\Delta x$ and phases $\Delta \phi$
appear as the result of their interaction[1-3].  For NSE the shifts 
$\Delta x$ and $\Delta \phi$ can be written as [1]:
\begin{eqnarray}
\Delta x_i = \zeta^{-1}_i \ln|Z_{1,2}| {~~~~}\Delta \phi_i= 2argZ_{1,2} ,
{~~~}i=1,2  \\						\nonumber
Z_{1,2} = \frac {2i(\zeta_1 + \zeta_2)-(v_1-v_2)} {2i(\zeta_1 - \zeta_2) -
(v_1-v_2)}.                                                         \label{f2}
\end{eqnarray}
Here $\zeta_i$  - is dimensionless parameter characterizing width
and amplitude of soliton, $v_i$ -is soliton velocity. 

To simplify the problem we will analyse the case of low density of soliton,
therefore we will consider the case when shift of soliton coordinate much more
then soliton width. In the region of parameters:

\begin{equation}
4(\zeta_1-\zeta_2)^2 < (v_1-v_2)^2 \ll 4(\zeta_1 + \zeta_2)^2, 
								  \label{f3}
\end{equation}
one can obtain that
\begin{equation}
\Delta x_1 \gg \zeta^{-1}_1, \Delta x_2 \gg \zeta^{-1}_2
                                                                   \label{f4}
\end{equation}
and
\begin{equation}
| \Delta \phi_1 | \approx | \Delta \phi_2 | \approx 1, 
							           \label{f5}
\end{equation}
that gives us the possibility to follow only for the evolution of soliton 
positions.

Let us introduce the distribution function of solitons $f(x,p,t)$.
To construct collision integral we follow [10] and apply Boltzmann 
approach taking into account that in contrast to usual particles solitons 
change there positions. Also it is necessary to consider that the 
interaction between solitons is absent when the distance become much more then 
soliton width. Let us analyse the number of solitons 
arriving to the point $(x,p_1)$ and leaving this point due to two soliton 
collisions for example in the case $p_2 >p_1>0$.

Due to following process:
\begin{eqnarray}
(x,p_1) \rightarrow  (x+ \Delta x_1,p_1)         \nonumber \\
(x,p_2) \rightarrow  (x- \Delta x_2,p_2)
								\label{f6}
\end{eqnarray}	
the number of solitons leaving the point $x,p_1$ can be written as:
\begin{eqnarray}
N^{(-)}=dx \int^{\infty}_{v_1} |v_1-v_2| f(x,p_1,t) f(x,p_2,t)dp_2. \label{f7}
\end{eqnarray}

The number of solitons arriving at the point $(x,p_1)$ and corresponding to 
following process
\begin{eqnarray}
(x-\Delta x_1,p_1) \rightarrow (x,p_1)              \nonumber  \\
(x-\Delta x_1,p_2) \rightarrow (x-\Delta x_1-\Delta x_2,p_2)
								\label{f8}
\end{eqnarray}
can be presented as:
\begin{equation}
N^{(+)}=dx\int^{\infty}_{v_1}|v_1-v_2| f(x-\Delta x_1,p_1,t) 
f(x-\Delta x_1,p_2,t) dp_2.
								\label{f9}
\end{equation}

From formulaes (7) and (9) one can write soliton-soliton collision integral 
in the case $p_2>p_1>0$. Considering all possible ratio for $p_1$ and $p_2$
soliton-soliton collision integral can be written as:
\begin{equation}
{\cal L}=\int_{-\infty}^{\infty} |v_1-v_2 | \{f(x-\Delta x_1,p_1,t)f(x-
\Delta x_1,p_2,t) - f(x,p_1,t)f(x,p_2,t)\} dp_2.                        
								\label{f10}
\end{equation}

In the approximation
\begin{equation}
|\Delta x_1| \approx |\Delta x_2| \approx |\Delta x|,{~~~}
\Delta x=\zeta^{-1} sign(v_1-v_2) \ln(\frac {4\zeta}{v_1-v_2})^2
								\label{f11}
\end{equation}
collision integral coincide with one obtained for Sine-Gordon equation in [10].
Assuming that $f(x,p,t)$ is slowly varying in scales comparable to 
$\Delta x$ and expanding $f(x,p,t)$ in powers of $\Delta x$ and keeping the 
leading terms we can rewrite  the expression  for ${\cal L}$ in the 
form:
\begin{equation}
{\cal L}=-\frac {\partial}{\partial x}  [f(x,p_1,t) u(x,p_1,t)] + 
\frac {1}{2} \frac {\partial^2}{\partial x^2} [f(x,p_1,t){\cal D}(x,p_1,t],
                                                                  \label{f12}
\end{equation}
where
\begin{equation} 
u(x,p_1,t)=\int_{-\infty}^{\infty} |v_1-v_2| \Delta x f(x,p_2,t) dp_2,
                                                                   \label{f13}
\end{equation}
describes the renormalization of soliton velocity and
\begin{equation}
{\cal D}(x,p_1,t)=\int_{-\infty}^{\infty} |v_1-v_2|(\Delta x)^2 f(x,p_2,t) 
dp_2,
                                                                   \label{f14}
\end{equation}
describes the solitons diffusion process.

The kinetic equation can be written as:
\begin{equation}
\frac {\partial f(x,p_1,t)}{\partial t}+ \frac {\partial}{\partial x}
[v_1 + u(x,p_1,t)]f(x,p_1,t)=\frac {1}{2} \frac {\partial^2}{\partial x^2}
{\cal D}(x,p_1,t)f(x,p_1,t).
                                                                \label{f15}
\end{equation}

From the kinetic equation (15) it is easy to show, that soliton-soliton 
collisions lead to the entropy production in the soliton gas in the case of 
nonuniform distribution function of soliton in coordinate space (see [10]). 
Applying standard methods from (10) it is easy to derive transport equations 
and calculate the relaxation time (see [10,11]).

3.Let us consider some applications that result from kinetic equation (15).
NSE appears in analysis of nonlinear electron plasma waves [12-15]. The 
conditions of solitons creation [16] and collapse [17,18] and interpretation of
Langmuir turbulence in soliton terms have been analysed in the frame of NSE 
(see [13-15]). 

On the base of proposed approach it is possible to estimate the relaxation 
time  $\tau_{ss}$for soliton gas. Indeed this estimation can be written as:
\begin{equation}
\frac {1}{\tau_{ss}} \sim (qx_0)^2 n v_T,                          \label{f16}
\end{equation}
where $x_0$ - is soliton size, $q$ is characteristic scale of inhomogeneity 
in soliton gas, $n$- is soliton density, $v_T= \sqrt {T /m}$ - is thermal 
velocity of solitons, $T$ -is temperature of solitons and $m$ is its mass.

Here
\begin{equation}
x_0=\zeta^{-1}\sqrt{\frac{\beta}{\alpha}},{~~}\beta=\frac{1}{2} 
\frac{dv_g}{dk},{~~}\alpha=-\frac{\partial^2 \omega}{\partial |\psi|^2}, 
                                                                  \label{f17}
\end{equation}
where $v_g$ - is group velocity of the high frequency wave, $\omega$ and $k$ - 
its frequency and wave vector,  $\psi$ -is slow varying nonlinear envelope.

To come nearer from proposed scheme to real situation it is necessary to take 
into account two facts: the interactions which destroy the integrability of 
the system are always existed and one dimensional soliton is unstable under 
the influence of 2D and 3D perturbations.
In the nonintegrable system solitons collide with momentum changing, but 
in the case close to completely integrable model it is not difficult to 
estimate the relaxation time $1/\tau^{\star}_{ss} \sim \Delta E/E$, where
$\Delta E/E$ is relative energy change due to collision. 
Therefore we have two steps of the relaxation: first one deal with shifts 
of the soliton positions and second deal with momentum exchange [10,19].
The characteristic time $\tau_{un}$ of soliton instability in 2D space
is finite and $\tau_{un} \sim L^2 $ (see [14]), where $L$ is characteristic 
size in perpendicular direction. Comparing these estimations with formula (16)
it is possible to conclude that above considered kinetic behaviour of solitons 
deal with shift of its positions can be realized as a 
intermediate regime before solitons collapse and can be interpreted as soliton
turbulence. Soliton turbulence phenomena have been considered in [15] were 
kinetic equation for many particle distribution function of soliton like waves
has been proposed. In fact in [15]  processes with momentum exchange only were 
considered.

4. Another application of NSE deal with soliton propagation in optic fibers 
[20,21]. For single mode case NSE can be written as: 
\begin{equation}
i\frac {\partial \chi}{\partial \xi}+ \frac{\partial^2 \chi}{\partial \tau^2}
+\sigma \chi |\chi|^2 = 0.                                      \label{f18}  
\end{equation}
Here 
$\chi$ is amplitude envelope of the pulse, $\xi = z/L_{\omega}$ - is the 
coordinate along the fiber, $L_{\omega}$ - is characteristic disperse length, 
$\tau = (z-v_0 t)/v_0 T_0$, where $v_0$  -is the group velocity, 
$T_0$ is the initial duration of the pulse, the value $\sigma$ is definited by
dispersion of group velocity and refraction coefficient. 

From applied point of view the problem of relaxation of solitons interacting 
with defects is one of the most actual among kinetic effects. The possible 
results of soliton - defect interaction are transmission, reflection or 
capturing of soliton by defect (see [22-25]). Dynamic properties of solitons 
in optic fiber interacting with dispersion-spectrum inhomogeneities have been 
studied in [26]. Obviously that the most natural way to analyze relaxation of 
solitons is to formulate soliton-defect collision integral. In general form 
it can be written as:
\begin{equation}
{\cal L}_{sd}=\int_{-\infty}^{\infty} W(p_1,p_2)\{f(x,p_1,t)-
f(x,p_2,t)\}dp_2.                
                                                                 \label{f19}
\end{equation}

Concrete expression for $W(p_1,p_2)$
in Sine-Gordon model was calculate in [11] for the case of elastic scattering 
of soliton by impurity. Here we estimate the relaxation time $\tau_{sd}$
in the case of low density of solitons comparing with defects concentration 
$C_{i}$. Really from dimensionally consideration:
\begin{equation}
\frac{1}{\tau_{sd}} \sim C_{i}\frac {\Delta E_{sd}}{E},
                                                                 \label{f20}
\end{equation}
where $\Delta E_{sd}$ is relative energy loss due to interaction with defect.

The defects appearance can deal with fiber irradiation. As a result the 
characteristics of light and sound propagation have changed under the influence
of the ionizing radiation. These effects have been applied  for measuring of 
the radiation dose [27,28]. Obviously that the relaxation time and kinetic 
coefficients of solitons depend from the intensity of fiber irradiation.

Let us consider the expression for the concentration of defects in the case
of $\gamma$-rays. The $\gamma$-rays can be generated in the processes of 
electron beam braking on heavy elements targets. 
The breaking radiation has continuous spectrum with the maximum energy of 
photons equal to the energy of electrons (see for example [29]). 
In this case [30]:
\begin{equation}
C_{i}=tJ_e K,
                                                                 \label{f21}
\end{equation}
where $t$ - is the time of irradiation, $J_e$ -is the current of electrons, 
coefficient $K$=2.1 for $Si$  [30] and energy of  $\gamma$-rays $E_{\gamma} 
\geq$20Mev.

Therefore:
\begin{equation}
J_e \sim \frac {1}{\tau_{sd} {~}t{~}(\Delta E_{sd}/E)}.
                                                                 \label{f22}
\end{equation}  

Consequently due to stability of soliton and low damping in optical fiber this
effect can be applied for measurement of irradiation dose.
	
This work can not been done without support of International Atomic Energy 
Agency, contract No 7996.
\newpage

				{\bf References}.
\vskip 0.25 in
1. Zakharov, V.E., Manakov, S.V., Novikov, S.P., Pitaevskii, L.P.: Theory of 
Solitons. Moscow: Nauka 1980 \\
2. Davydov, A.S.: Solitons in Molecular Systems 2nd edn. Kiev:Naukova Dumka 
1988 \\
3. Kosevich, A.M., Kovalev, A.S.: Introduction in Nonlinear Mechanics. Kiev:
Naukova Dumka 1989 \\
4. Zakharov, V.E.: Zh.Eksp.Theor.Fiz. 60, 993 (1971) \\
5. Wada, Y., Schrieffer, J. R.: Phys.Rev. B 18, 3897 (1978) \\
6. Theodorakopoulos, N.: Z.Phys. B 33, 385 (1979)\\
7. Fesser, K.:Z.Phys. B 39, 47 (1979)\\
8. Baryakhtar, V.G., Baryakhtar, I.V., Ivanov, B.A., Sukstanskii A.L.: In: 
Proc. 2nd Int. Symp. on  Selected Topics in Statistical Mechanics p.417. 
Dubna:JIRN 1981 \\
9. Malomed, B.A., Flytzanis N.: Phys. Rev. E  48, R5 (1993) \\
10. Baryakhtar,I.V., Baryakhtar, V.G., Economou E.N.: Phys.Lett. A 207, 67 
(1995) \\
11. Baryakhtar, I.V., Baryakhtar, V.G., Economou E.N. (to appear in 
Europhys. Lett) \\ 
12. Chen, F.F.: Introduction to Plasma Physics and Controlled Fusion. 
New York: 
Plenum Press 1984 \\ 
13. Foundation of Plasma Physics, vol.2 Galeev, A.A., Sudan, R. (eds) Moscow: 
Energoatomizdat 1984 \\
14. Kadomtsev, B.B. Collective Phenomena in Plasma 2nd edn. Moscow: Nauka 
1988 \\
15. Sagdeev R.Z., Zaslavskii G.M.: Introduction to Nonlinear Physics. Moscow:
Nauka 1988 \\
16. Vedenov, A.A., Rudakov L.I.: Doklady 159, 2059 (1964) \\
17.Thornhill, S.G., Ter-Haar,D.: Phys.Reports 43C,43 (1972)\\
18. Zakharov, V.E., Zh.Eksp.Theor.Fis. 62, 1995 (1972) \\
19. Baryakhtar I.V. In: First Meeting Report of the CRP on "Development of 
Plasma  Heating and Diagnostic Systems in Institutes in Developing Countries
using middle- and small- scale plasma devices". Appendix A8,Vienna IAEA 1996 \\
20. Hasegawa, A., Tappert F.: Appl. Phys. Lett. 23, 142  (1973)\\
21. Sisakyan, I.N., Shvartsburg, A.B.: Quantum Electronics 11, 1703 (1984)\\
22. Bass, F.G., Kivshar, Yu.S., Konotop V.V., Sinitsin Yu.A.: Phys. Reports 
157, 63 (1988) \\
23. Li Qiming, Pnevmatikos St., Economou E.N., Soukoulis C.M.: Phys.Rev. B 37, 
3534 (1988)\\
24. Li Qiming, Soukoulis C.M., Pnevmatikos St., Economou E.N. : Phys.Rev. B 38,
11888 (1989)\\
25. Fraggis, T., Pnevmatikos St., Economou E.N.: Phys. Lett A 142, 361 
(1989)\\
26. Vysloukh, V.A., Serkin, V.N., Danilenko, A.Yu., Samarina E.V.: Quantum 
Electronics 22, 1129 (1995)\\
27. Vassilopoulos, C., Kortis, A., Mantakas, C.:IEE PROCEEDINGS-J 140, 267 
(1993)\\
28. Poret, J.C., E.Lindgren, Rosen M., Suter J.J., Rifkind J.M.:
J.Non-Cryst. Solids 160, 82 (1993)\\
29. Kovalev, V.P.:  Secondary Irradiation of Electron Accelerator. Moscow: 
Energoatomizdat 1979\\
30. Zablotskii,V.V., Ivanov I.A.Fiz. Tech. Poluprovod 20, 625 (1986)
\end{document}